\documentclass{IEEEtran}
\usepackage{cite}
\ifCLASSINFOpdf
  \usepackage[pdftex]{graphicx}
  \graphicspath{{./gfx/}}
\else
  \usepackage[dvips]{graphicx}
  \graphicspath{{./gfx/}}
  \DeclareGraphicsExtensions{.eps}
\fi
\usepackage[cmex10]{amsmath}
\usepackage{amssymb, amsfonts, amsthm}

\usepackage{array}
\usepackage[tight,footnotesize]{subfigure}
\usepackage{url}
\newcommand{\ignore}[1]{}
\renewcommand{\Pr}{\mathbb{P}} % probability
\newcommand{\T}{T} % stopping time
\DeclareMathOperator{\EV}{\mathbb{E}} % expected value
\DeclareMathOperator{\Var}{\mathbb{V}\!ar} % variance

\DeclareMathOperator{\LLR}{\mathcal{L}}
\DeclareMathOperator{\LR}{\Lambda}

\DeclareMathOperator{\ARL}{ARL}
\DeclareMathOperator{\SADD}{SADD}

\DeclareMathOperator{\STADD}{STADD}

\renewcommand{\le}{\leqslant} % AMS le ge
\renewcommand{\ge}{\geqslant}
\renewcommand{\sim}{\thicksim}

\begin{document}

\title{Efficient Computer Network Anomaly Detection by Changepoint Detection Methods}

\author{\IEEEauthorblockN{Alexander G. Tartakovsky,~\IEEEmembership{Senior Member,~IEEE}, Aleksey S. Polunchenko\IEEEauthorrefmark{1} and Grigory Sokolov}\\
\IEEEauthorblockA{Department of Mathematics
and
Center for Applied Mathematical Sciences,
University of Southern California,
3620 S. Vermont Ave., KAP 108,
Los Angeles, CA 90089-2532,
United States of America\\
Email: \{tartakov,polunche,gsokolov\}@usc.edu}\\
\IEEEauthorblockA{\IEEEauthorrefmark{1}Currently with the Department of Mathematical Sciences, State University of New York (SUNY) at Binghamton, Binghamton, NY 13902-6000, United States of America\\
Email: aleksey@binghamton.edu}}



\maketitle

%-------------------------------------------------------------------------------------------------%
\begin{abstract}
%\boldmath
We consider the problem of efficient on-line anomaly detection in computer network traffic. The problem is approached statistically, as that of sequential (quickest) changepoint detection. A {\em multi-cyclic setting} of quickest change detection is a natural fit for this problem. We propose a novel score-based multi-cyclic detection algorithm. The algorithm is based on the so-called Shiryaev--Roberts procedure. This procedure is as easy to employ in practice and as computationally inexpensive as the popular Cumulative Sum chart and the Exponentially Weighted Moving Average scheme. The likelihood ratio based Shiryaev--Roberts procedure has appealing optimality properties, particularly it is {\em exactly} optimal in a multi-cyclic setting geared to detect a change occurring at a far time horizon. It is therefore expected that an intrusion detection algorithm based on the Shiryaev--Roberts procedure will perform better than other detection schemes. This is confirmed experimentally for real traces. We also discuss the possibility of complementing our anomaly detection algorithm with a spectral-signature intrusion detection system with false alarm filtering and true attack confirmation capability, so as to obtain a synergistic system.
\end{abstract}


% \fi
\IEEEpeerreviewmaketitle

%-------------------------------------------------------------------------------------------------%
\section{Introduction}
\label{sec:intro}

The Internet has never been a safe place and designing automated and efficient techniques for rapid detection of computer network anomalies  (e.g., due to intrusions) never ceased to be a topical problem in cybersecurity~\cite{Ellis+Speed:Book01}. Many existing anomaly-based Intrusion Detection Systems (IDS-s) operate by applying the machinery of statistics to comb through the passing traffic looking for a deviation from the traffic's normal profile~\cite{Thatte+etal:IEEE-TN11,Thatte+etal:IEEE-GIS08,Tartakovsky+etal:SP06,Tartakovsky+etal:SM06+discussion,Tartakovsky+etal:JSM05}. By way of example, the Sequential Probability Ratio Test (SPRT)~\cite{Wald:Book47}, the Cumulative Sum (CUSUM) chart~\cite{Page:B54}, and the Exponentially Weighted Moving Average (EWMA) inspection scheme~\cite{Roberts:T59} are the {\it de facto} ``workhorse'' of the community. The CUSUM and EWMA methods  come from the area of sequential changepoint detection, a branch of statistics concerned with the design and analysis of a {\em fastest} way to detect a change (i.e., an anomaly) in the state of a phenomenon (time process) of interest~\cite{Poor+Hadjiliadis:Book08,Basseville+Nikiforov:Book93}.

Yet another changepoint detector popular in statistics is the Shiryaev--Roberts (SR) procedure~\cite{Shiryaev:SMD61,Shiryaev:TPA63,Roberts:T66}. Though practically unknown in the cybersecurity community, the SR procedure is as computationally simple as the CUSUM chart or the EWMA scheme. However, unlike the latter two, the SR procedure is also the best one can do (i.e., exactly optimal) in a certain multi-cyclic setting~\cite{Pollak+Tartakovsky:SS09}, a natural fit in the computer network anomaly detection context. The aim of this work is to offer a novel multi-cyclic anomaly detector using the SR procedure as the prototype. Due to the exact multi-cyclic optimality of the SR procedure, the proposed algorithm is expected to outperform other detection schemes, in particular the multi-cyclic CUSUM procedure. We confirm this experimentally using real data.

The remainder of the paper is organized as follows. Section~\ref{sec:change-point} provides an introduction to the subject of changepoint detection. In Section~\ref{sec:trans}, we present our anomaly detection algorithm. In Section~\ref{sec:case-study}, we illustrate our algorithm at work. In Section~\ref{sec:further-discussion}, we comment on how to improve the performance of the algorithm. Lastly, Section~\ref{sec:conclusion} draws the conclusions.

%-------------------------------------------------------------------------------------------------%
\section{Quickest Changepoint Detection}
\label{sec:change-point}

Quickest changepoint detection is a study of techniques to detect a change (``disorder'') in the state of a time process, usually from ``normal'' to ``abnormal''; inference about the process' current state is made from a series of quantitative random observations (e.g., measurements corrupted by noise). The sequential setting assumes the series is amassed one at a time, and so long as the recorded data behavior suggests the process is in its ``normal'' state it is let to continue. However, if the observations hint that the process' state may have switched to ``abnormal'', one's aim is to detect the true change as quickly as possible for a given risk associated with false alarms, so that an appropriate response can be provided in a timely manner. The time instance at which the state of the process changes is referred to as the {\em changepoint}, and the challenge is that it is not known in advance. This is known as the {\em sequential (quickest) changepoint detection problem}. For lack of space, from now on we will focus only on the basic {\em iid} version of this problem; a general non-iid case is surveyed, e.g., in~\cite{Tartakovsky+Moustakides:SA10, Polunchenko+Tartakovsky:MCAP11}.

Suppose one is able to sequentially collect a series of independent random observations, $\{X_n\}_{n\ge1}$, such that $X_1,\ldots,X_\nu$ are each distributed according to a known probability density function (pdf) $f$, while $X_{\nu+1},X_{\nu+2},\ldots$ each adhere to a pdf $g\not\equiv f$, also known. The time index $\nu$ (i.e., the changepoint) is assumed unknown non-random number; for cases that regard $\nu$ as a random variable, see, e.g.,~\cite{Shiryaev:SMD61,Shiryaev:TPA63}. One's aim is to detect that the observations' common distribution has changed. The challenge is to do so with as few observations as possible following the changepoint, subject to a tolerable limit on the risk of making a false detection.

Statistically, the problem is to sequentially differentiate between the hypotheses $\mathcal{H}_k\colon\nu=k$, $0\le k<\infty$ (i.e., that the data $\{X_n\}_{n\ge1}$ change their statistical profile at time instance $\nu=k$, $0\le k<\infty$) and $\mathcal{H}_\infty\colon\nu=\infty$ (i.e., that no change ever occurs). To test $\mathcal{H}_k$ against $\mathcal{H}_\infty$ one first constructs the corresponding likelihood ratio, which for the iid scenario has the form
\begin{align*}
\LR_{k:n}= \prod_{j=k+1}^n\LR_j,\;\;\text{where}\;\;\LR_j=\frac{g(X_j)}{f(X_j)},
\end{align*}
and it is understood that $\LR_{k:n}\equiv 1$ for $k\ge n$.

Next, as each new observation becomes available to test the hypotheses, the sequence $\{\LR_{k:n}\}_{1\le k\le n}$ is turned into a {\em detection statistic}. To this end, one can either use the maximum likelihood principle or the (generalized) Bayesian
approach. In the former case the corresponding detection statistic is
\begin{equation}\label{eq:CUSUM_V}
V_n=\max_{1\le k\le n}\LR_{k:n}, \quad n\ge1,
\end{equation}
i.e., the famous CUSUM statistic. The Bayesian statistic depends on the changepoint's prior distribution. As in our case the changepoint, $\nu$, is assumed unknown, the corresponding quasi-Bayesian (or generalized Bayesian) detection statistic can be defined as
\[
    R_n=\sum_{k=1}^n\LR_{k:n}, \quad n\ge1.
\]
One can view $\{R_n\}_{n\ge1}$ as being the average of the sequence $\{\LR_{k:n}\}_{1\le k\le n}$ with respect to an (improper) uniform prior distribution imposed on $\nu$; see, e.g.,~\cite{Girschick+Rubin:AMS52,Shiryaev:SMD61,Shiryaev:TPA63,Tartakovsky+Moustakides:SA10,Polunchenko+Tartakovsky:MCAP11}.

Once the detection statistic is chosen, it is supplied to an appropriate sequential detection procedure. A detection procedure is a stopping time, $\T$, which is a function of the observed data, $\{X_n\}_{n\ge1}$. The meaning of $\T$ is that after observing $X_1,\ldots,X_T$ it is declared that the change is in effect. That may or may not be the case. If it is not, then $T\le\nu$, and it is said that a false alarm has been sounded.

Henceforth, let $\Pr_\nu(\cdot)$ and $\Pr_\infty(\cdot)$ denote the probability measures, respectively, when the change occurs at time instant $0\le\nu<\infty$, and when no change ever occurs. Likewise, let $\EV_\nu[\cdot]$ and $\EV_\infty[\cdot]$ be the corresponding expectations.

Lorden~\cite{Lorden:AMS71} suggested to measure the risk of raising a false alarm via the Average Run Length (ARL) to false alarm $\ARL(\T)=\EV_\infty[\T]$ and showed that the CUSUM procedure has certain minimax properties in the class of detection procedures
\begin{align*}
\Delta(\gamma)
&=
\Bigl\{\T\colon\ARL(\T)\ge\gamma\Bigr\}
\end{align*}
for which the ARL to false alarm is no ``worse'' than the desired {\it a~priori} chosen level $\gamma>1$. See also Moustakides~\cite{Moustakides:AS86} who proved that CUSUM is in fact strictly minimax with respect to Lorden's criterion for every $\gamma>1$.

A practically appealing way to measure the detection speed is Pollak's~\cite{Pollak:AS85} ``worst-case'' (Supremum) Average Delay to Detection (ADD), conditional on a false alarm not having been previously sounded, i.e.,
\[
    \SADD(\T)=\max_{0\le k<\infty} \EV_k[\T-k|\T>k].
\]
The minimax quickest changepoint detection problem is to find $\T_{\mathrm{opt}}\in\Delta(\gamma)$
such that
\[
\SADD(\T_{\mathrm{opt}})=\inf_{T\in\Delta(\gamma)}\SADD(\T) \quad \text{for all}~ \gamma>1.
\]
To date, this problem remains open, and only asymptotic (as $\gamma\to\infty$) solutions have been obtained~\cite{Pollak:AS85,Tartakovskyetal-TPA12}.

The CUSUM chart~\cite{Page:B54} has been popular in many areas of engineering and computer science, including cybersecurity. It iteratively maximizes the log-likelihood ratio (LLR) with respect to the changepoint $\nu$, and stops once the maximum exceeds a certain threshold. More specifically, the CUSUM procedure is based on the statistic $W_n=\max\{0, \log V_n\}$, where $V_n$ is defined in \eqref{eq:CUSUM_V}, which is computed recursively
\[
    W_n=\max\{0,W_{n-1}+\LLR_n\}, \quad n\ge1, ~~ W_0=0 .
\]
Here $\LLR_n=\log\LR_n$ is the LLR. The corresponding stopping rule is
\[
    \mathcal{C}_h=\min\{n\ge1\colon W_n\ge h\},
\]
where $h>0$ is a detection threshold preset so as to achieve the desired level of false alarms $\gamma>1$, and thus guarantee that $\mathcal{C}_h\in\Delta(\gamma)$. This can be achieved by setting $h=h_\gamma\ge\log\gamma$, since $\ARL(\mathcal{C}_h)\ge e^h$ for any $h>0$~\cite{Lorden:AMS71}. For large values of $\gamma$ more ``careful'' selection of $h$ is possible~\cite{Polunchenko+Tartakovsky:MCAP11}.

\begin{figure*}[!t]
    \centering
    \subfigure[An example of the behavior of a process of interest with a change in mean at time $\nu$.]{
        \includegraphics[width=0.9\textwidth]{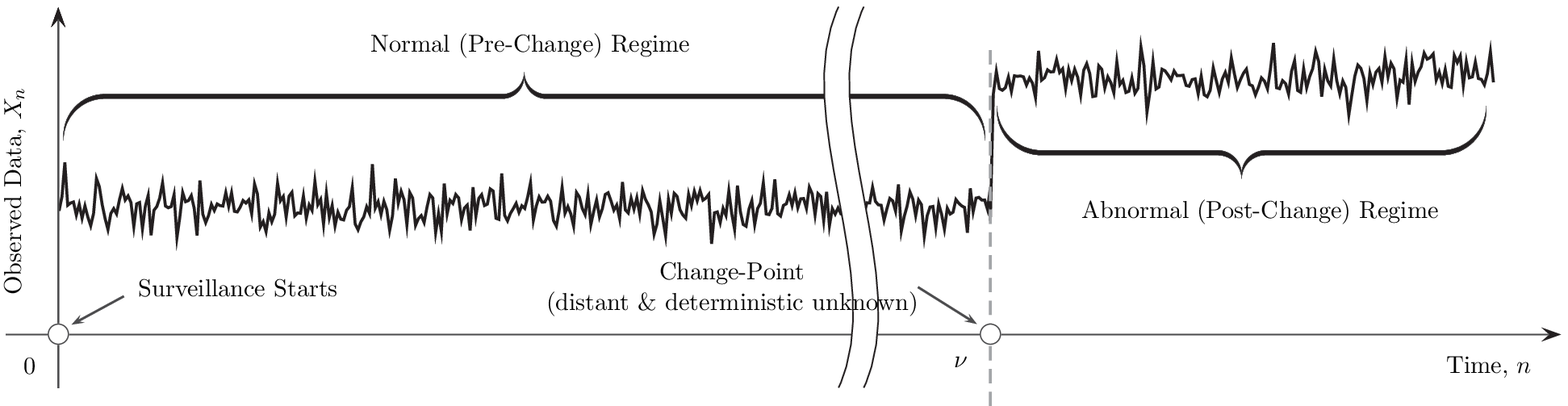}
    } % /subfigure
    \subfigure[Typical behavior of the detection statistic in the multi-cyclic mode.]{
        \includegraphics[width=0.9\textwidth]{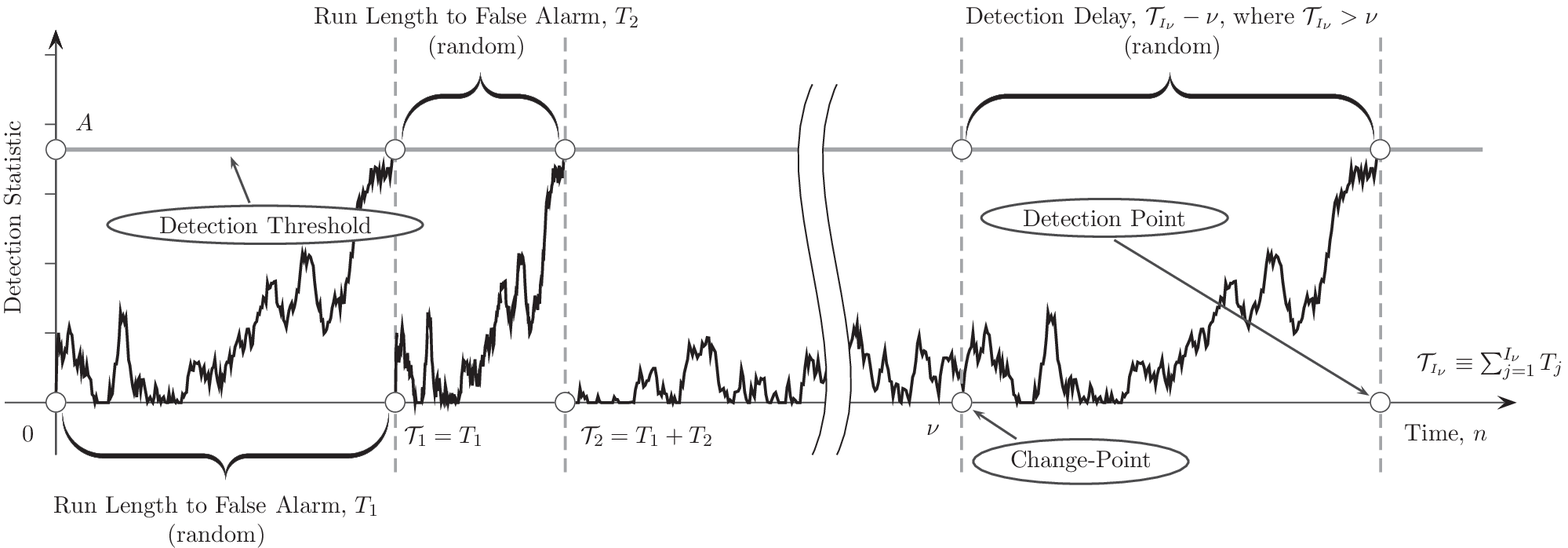}
    } % /subfigure
    % figure caption to be below the figure
    \caption{Multi-cyclic changepoint detection in a stationary regime.}
    \label{fig:multi-cyclic-idea}
\end{figure*}

Consider now a context in which it is of utmost importance to detect the change as quickly as possible, even at the expense of raising many false alarms (using a repeated application of the same stopping rule) before the change occurs. Put otherwise, in exchange for the assurance that the change will be detected with maximal speed, we agree to go through a ``storm'' of false alarms along the way (the false alarms are ensued from repeatedly applying the same detection rule, starting from scratch after each false alarm). This scenario is shown in Figure~\ref{fig:multi-cyclic-idea}.

Formally, let $T_1,T_2,\ldots$ be sequential independent repetitions of the stopping time $\T$, and let ${\cal T}_j=T_1+T_2+\cdots+T_j$, $j\ge1$, be the time of the $j$-th alarm. Define $I_\nu=\min\{j\ge1\colon {\cal T}_j>\nu\}$. In other words, ${\cal T}_{\scriptscriptstyle I_\nu}$ is the time of detection of a true change that occurs at $\nu$ after $I_\nu-1$ false alarms have been raised. Write
\begin{align*}
\STADD(\T)
&=
\lim_{\nu\to\infty}\EV_\nu[{\cal T}_{\scriptscriptstyle I_\nu}-\nu]
\end{align*}
for the limiting value of the average delay to detection referred to as the {\em Stationary Average Delay to Detection} (STADD).
The multi-cyclic changepoint detection problem is to find $\T_{\mathrm{opt}}\in\Delta(\gamma)$ such that
\[
\STADD(\T_{\mathrm{opt}})=\inf_{\T\in\Delta(\gamma)}\STADD(\T) \quad \text{for every}~\gamma>1.
\]
This formulation is instrumental in detecting a change that takes place in a distant future (i.e., $\nu$ is large), and is preceded by a stationary flow of false detections. Such scenarios are a commonplace in the area of computer network anomaly detection.

As has been shown by Pollak and Tartakovsky~\cite{Pollak+Tartakovsky:SS09},  the so-called Shiryaev--Roberts (SR) procedure~\cite{Shiryaev:TPA63,Roberts:T66} is {\em exactly} optimal for every $\gamma >1$ with respect to the stationary average detection delay $\STADD(\T)$. Thus, in the multi-cyclic setting the SR procedure is a better alternative to the popular CUSUM and EWMA schemes.

The SR rule stops at time instance
\[
    \mathcal{S}_A=\min\{n\ge1\colon R_n\ge A\},
\]
where the SR statistic is given by the recursion
\[
    R_n=(1+R_{n-1})\LR_n, \quad n\ge1, \quad R_0=0.
\]
Here $A>0$ is a detection threshold set {\it a~priori} so as to ensure $\mathcal{S}_A\in\Delta(\gamma)$ for a desired $\gamma>1$. It can be easily shown~\cite{Pollak:AS87} that $\ARL(\mathcal{S}_A)\ge A$ for all $A>0$, so choosing the detection threshold as $A_\gamma=\gamma$ will guarantee $\mathcal{S}_{A}\in\Delta(\gamma)$. A very accurate asymptotic approximation  $\ARL(\mathcal{S}_A)\sim A/v$, $A\to \infty$ is also possible, where $0<v<1$ is a constant which is a subject of renewal theory. See, e.g.,  \cite{Pollak:AS87}.

%-------------------------------------------------------------------------------------------------%
\section{Transition to Cybersecurity}
\label{sec:trans}

The above somewhat abstract introduction to sequential changepoint detection is straightforward to put in the context of anomaly detection in computer network traffic. As network anomalies typically take place at {\em unknown} points in time and entail changes in the traffic's statistical properties, it is intuitively appealing to formulate the problem of computer network anomaly detection as that of a quickest changepoint detection: to detect changes in the statistical profile of network traffic  as rapidly as possible, while maintaining a tolerable level of the risk of making a false detection.

\begin{figure}[h]
    \centering
    \subfigure[Observed data.]{
        \includegraphics[width=0.475\textwidth]{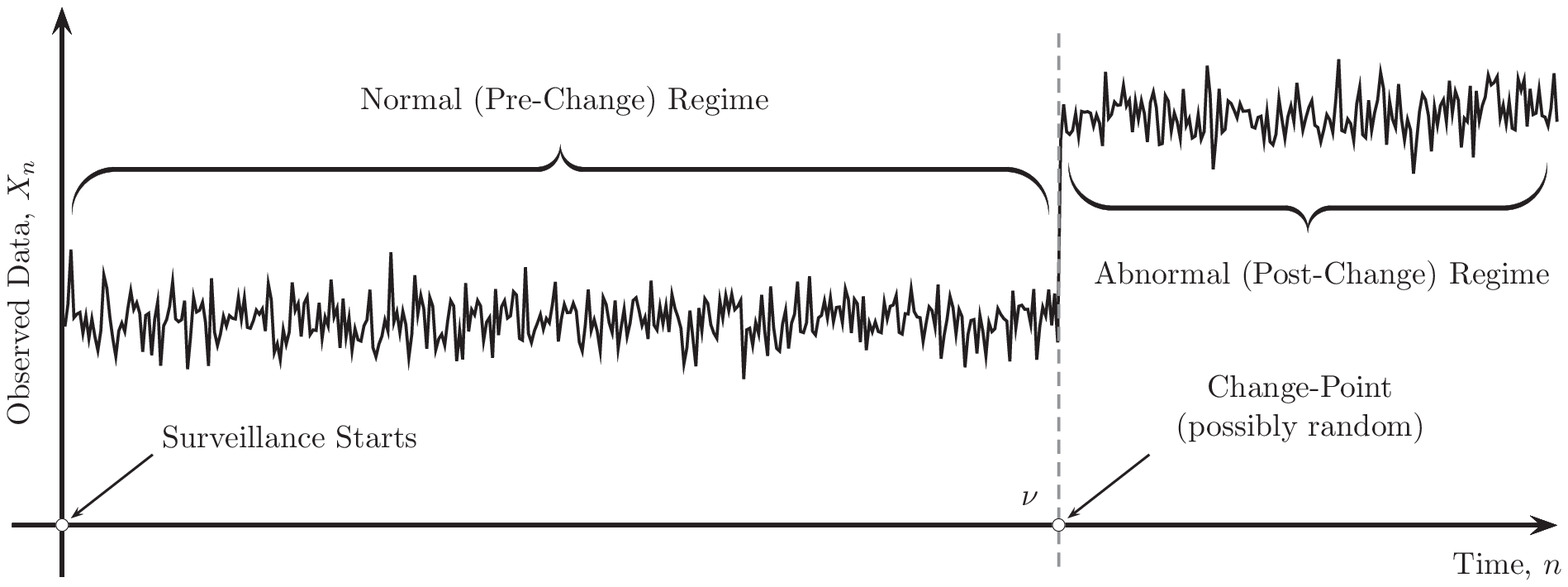}
    }
    \subfigure[Two possible terminal decisions: either a false alarm (dashed), or a correct but delayed detection (solid).]{
        \includegraphics[width=0.475\textwidth]{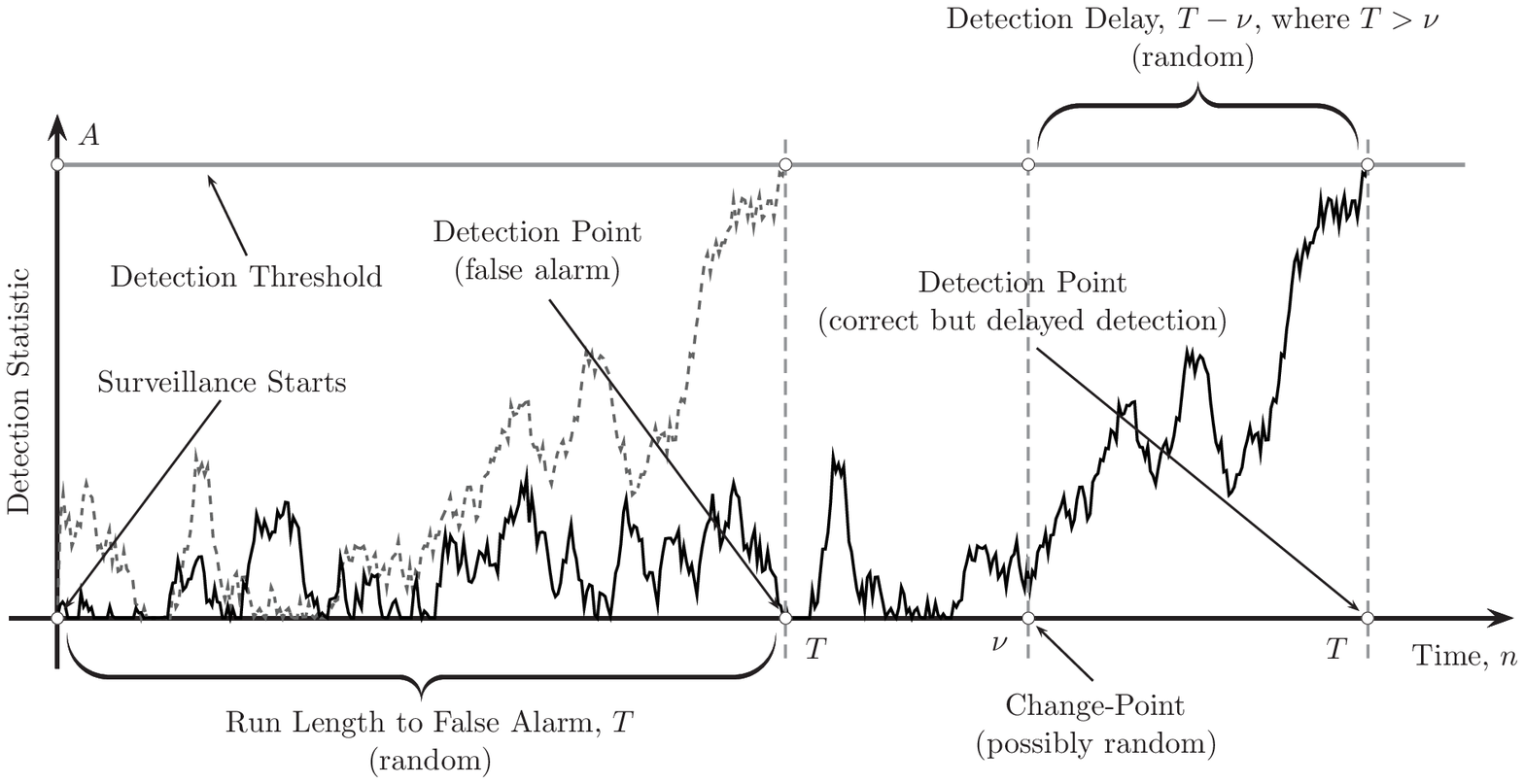}
    }
    \caption{Typical run of the detection statistic.}
    \label{fig:Statistics}
\end{figure}

It is common that in practice neither pre- nor post-anomaly distributions are known. As a result, traffic's pre- and post-anomaly profile is poorly understood, and one can no longer rely on the likelihood ratios. Hence, an alternative approach is required. Let us first consider a typical behavior of the CUSUM and SR statistics. As long as the observed sequence $\{X_n\}_{n\ge1}$ is in the normal mode, the detection statistics $\{W_n\}_{n\ge1}$ and $\{R_n\}_{n \ge 1}$ behave as if they were ``afraid'' of approaching the detection thresholds $h$ and $A$ respectively (although it is still possible that the thresholds will be crossed, in which case a false alarm will be raised). However, as soon as $X_{\nu+1}$ -- the first data point affected by an anomaly -- is recorded, the behavior of $W_n$ and $R_n$ changes completely, so that they now eagerly try to hit the thresholds. Formally, this means that $\EV_\infty[\LLR_n]<0$ and $\EV_\nu[\LLR_n]>0$, $\nu < n$. That is, the detection statistic has a negative drift under the normal regime, and a positive drift in an anomaly situation. A typical behavior of the detection statistic in pre- and post-change regimes is shown in Figure~\ref{fig:Statistics}.

Consider now the following score-based modification of the SR procedure
\[
\tilde{R}_n=(1+\tilde{R}_{n-1})e^{S_n}, \quad n\ge1, \quad \tilde R_0=0
\]
with the corresponding stopping time being
\[
\tilde{\mathcal{S}}_{A}=\min\{n\ge1\colon\tilde{R}_n\ge A\},
\]
where $A>0$ is an {\it a~priori} chosen detection threshold. Similarly for CUSUM,
\[
    \tilde{W}_n=\max\{0,\tilde{W}_{n-1} + S_n\}, \quad n\ge1, \quad \tilde{W}_n=0
\]
with the corresponding stopping time being
\[
\tilde{\mathcal{C}}_{h}=\min\{n\ge1\colon\tilde{W}_n\ge h\}, \quad h \ge 0.
\]
Here $S_n(X_1,\ldots,X_n)$ are the selected score functions. Clearly, so long as
\[
\EV_\infty[S_n(X_1,\ldots,X_n)]<0~~\text{and}~~\EV_\nu[S_n(X_1,\ldots,X_n)]>0,
\]
for all $\nu \ge 0$, the SR and CUSUM detection procedures designed using such score functions in place of the likelihood ratio will work, though they will not be optimal anymore. Their behavior will be similar to that shown in  Figure~\ref{fig:Statistics}. Score functions $S_n$ can be chosen in a number of ways and each particular choice depends crucially on the expected type of change. In the applications of interest, the detection problem can be usually reduced to detecting changes in mean values along with variances (mean and variance shifts).

Let
\[
\mu_\infty=\EV_\infty[X_n], ~~ \sigma_\infty^2=\Var_\infty[X_n]
\]
and
\[
\mu=\EV_0[X_n], ~~ \sigma^2=\Var_0[X_n]
\]
denote the pre- and
post-anomaly mean values and variances, respectively. Write $Y_n=(X_n-\mu_\infty)/\sigma_\infty$ for the centered and scaled observation at time $n$. In the real-world applications the pre-change parameters $\mu_\infty$ and $\sigma_\infty^2$ are estimated from the training data and periodically re-estimated due to the non-stationarity of network traffic at large time-scales. We suggest the score $S_n$ of the linear-quadratic form
\begin{equation}\label{eq:score}
S_n(Y_n) = C_1 Y_n + C_2 Y_n^2 - C_3,
\end{equation}
where $C_1$, $C_2$ and $C_3$ are positive design numbers assuming for concreteness that the change leads to an increase in both mean and variance. In the case where the variance either does not change or changes relatively  insignificantly compared to the change in mean, the coefficient $C_2$ may be set to zero. In the opposite case where the mean changes only slightly compared to the variance, we take $C_1=0$. The first case appears to be typical for many cybersecurity applications, for example for ICMP and UDP Denial-of-Service (DoS) attacks (see~\cite{Tartakovsky+etal:SP06, Tartakovsky+etal:SM06+discussion} where the linear score-based CUSUM has been proposed). However, in certain cases, such as the one considered below in Section~\ref{sec:case-study}, both the mean and variance change quite significantly.

Note that the score given by \eqref{eq:score} with
\begin{equation}\label{eq:design_C}
  C_1 = \delta q^2,  \quad  C_2 = (1 - q^2) / 2,   \quad C_3 = \delta^2 q^2 / 2 - \log q,
\end{equation}
where $q=\sigma_\infty/\sigma$, $\delta=(\mu-\mu_\infty)/\sigma_\infty$, is optimal if pre- and post-change distributions are Gaussian with known putative values $\mu$ and $\sigma^2$. This is true because in the latter case $S_n$ is the log-likelihood ratio. If one believes in the Gaussian model (which sometimes is the case), then selecting $q=q_0$ and $\delta=\delta_0$ with some design values $q_0$ and $\delta_0$ provides reasonable operating characteristics for $q < q_0$ and $\delta > \delta_0$ and optimal characteristics for $q=q_0$ and $\delta=\delta_0$. However, it is important to emphasize  that the proposed score-based SR procedure does not assume that the observations have Gaussian pre- and post-change distributions.

Further improvement may be achieved by using either mixtures or adaptive versions with generalized likelihood ratio-type statistics \cite{Pollak:AS87, Lorden:AMS71}.

Based on the previous reasoning (see Section~\ref{sec:change-point}) we expect the multi-cyclic score-based SR procedure to perform better than the analogous CUSUM procedure.

%-------------------------------------------------------------------------------------------------%
\section{A Case Study}
\label{sec:case-study}

\begin{figure}[h!]
    \centering
    \includegraphics[width=0.4\textwidth]{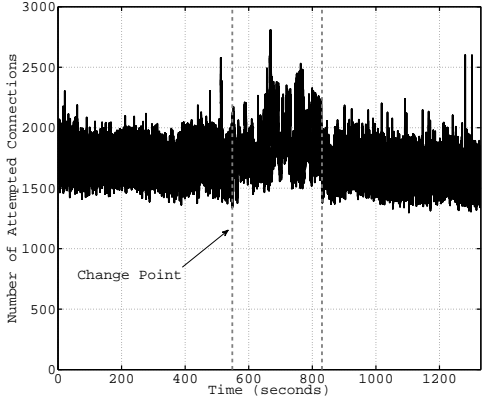}
    \caption{SYN flood attack: number of attempted connections.}
    \label{fig:SYN}
\end{figure}

We now present the results of testing the proposed detection algorithms on a real Distributed DoS (DDoS) attack, namely, SYN flood attack. The aim of this attack is to congest the victim's link with a series of SYN requests so as to have the victim's machine exhaust all of its resources and stop responding to legitimate traffic. This kind of an attack clearly creates a volume-type anomaly in the victim's traffic flow. The data is courtesy of the Los Angeles Network Data Exchange and Repository (LANDER) project (see~\url{http://www.isi.edu/ant/lander}). Specifically, the trace is flow data captured by Merit Network Inc. (see~\url{http://www.merit.edu}). The attack is on a University of Michigan IRC server. It starts at roughly $550$ seconds into the trace and has a duration of $10$ minutes. The attacked IP is anonymized to 141.213.238.0. Figure~\ref{fig:SYN} shows the number of attempted connections or the connections birth rate as a function of time. While the attack can be seen to the naked eye, it is not completely clear when it starts. In fact, there is a spike in the data (fluctuation) before the attack. Also, controlling the false alarm rate with an automatic detection system is a challenge.

We used the number of connections during 20 msec batches as the observations $X_n$. We estimated the connections birth rate average and variance for legitimate traffic and for attack traffic; in both cases, to estimate the average we used the usual sample mean, and to estimate the variance we used the usual sample variance. For legitimate traffic, the average is about $\mu_\infty=1669.09$ connections per 20 msec, and the standard deviation is in the neighborhood of $\sigma_\infty=113.884$ connections per 20 msec. For attack traffic, the numbers are $\mu=1887.56$ and $\sigma=218.107$, respectively. We can now see the effect of the attack: it leads to a considerable increase in the mean and standard deviation of the connections birth rate.

\begin{figure}[!h]
    \centering
    \subfigure[Legitimate (pre-attack) traffic.]{
        \includegraphics[width=0.4\textwidth]{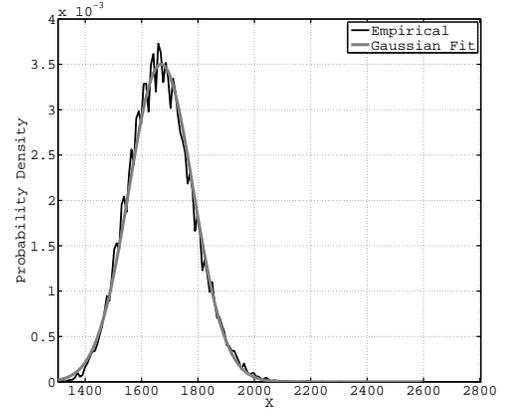}
        \label{fig:SYN-pre-change-pdf}
    }
    \subfigure[Attack traffic.]{
        \includegraphics[width=0.4\textwidth]{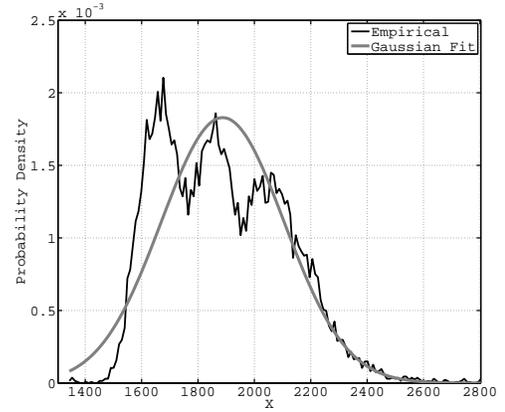}
        \label{fig:SYN-post-change-pdf}
    }
    \caption{SYN flood attack: connections birth rate pdf with a Gaussian fit.}
    \label{fig:SYN-pdf}
\end{figure}

\ignore{
It is interesting to verify whether traffic expressed via the number of connections may be approximated by the normal distribution.  To see how well the data agree with this hypothesis, we now access the goodness of fit of a Gaussian distribution and the data. Figure~\ref{fig:SYN-pdf} shows the empirical densities of the connections birth rate for legitimate and attack traffic. Note the similarity to the Gaussian distribution for legitimate traffic. However, for attack traffic, the distribution is not as close to Gaussian.
} % End of ignore
We now perform a basic statistical analysis of the connections birth rate distribution. Figure~\ref{fig:SYN-pdf} shows the empirical densities of the connections birth rate for legitimate and attack traffic. It so happens that for given data, legitimate traffic appears to resemble the Gaussian process. However, for attack traffic, the distribution is not as close to Gaussian.
\ignore{

The same conclusion can be made from an inspection of the corresponding Q-Q plots (quantile-quantile) shown in Figure~\ref{fig:SYN-QQ}. Specifically, the Q-Q plot for legitimate traffic is shown in Figure~\ref{fig:SYN-pre-change-qq} and  for attack traffic in Figure~\ref{fig:SYN-post-change-qq}. Both plots are for centered and scaled data, so the fitted Gaussian distribution is the standard normal distribution. The fact that the plot for legitimate traffic is a straight line also confirms the ``Gaussianness'' of the distribution. For attack traffic, we see that the distribution differs from the standard Gaussian.
\begin{figure}[!htb]
    \centering
    \subfigure[Legitimate (pre-attack) traffic.]{
        \includegraphics[width=0.4\textwidth]{TCP_0_Inf_Base16_QQ_Benign}
        \label{fig:SYN-pre-change-qq}
    }
    \subfigure[Attack traffic.]{
        \includegraphics[width=0.4\textwidth]{TCP_0_Inf_Base16_QQ_Attack}
        \label{fig:SYN-post-change-qq}
    }
    \caption{SYN flood attack: Q-Q plots for connections birth rate distribution vs. standard Gaussian distribution.}
    \label{fig:SYN-QQ}
\end{figure}
} % End of ignore
\ignore{
We have implemented the score-based multi-cyclic SR and CUSUM procedures with the linear-quadratic score \eqref{eq:score}.  When choosing design parameters we assume the Gaussian model for pre-attack traffic, which agrees with fitting presented above. {\bf How exactly did you use this assumption to estimate design parameters and which ones?? This is exactly what he did not like.} We pick thresholds $A \approx 1.9 \cdot 10^3, h \approx 6.68$ to get the same level of $\ARL \approx 500$ samples (i.e., $10$ sec) for both procedures, which is estimated using Monte Carlo simulations under existing pre-change distribution. Parameters $C_1, C_2$, and $C_3$ are chosen according to formulas \eqref{eq:design_C} with $q_0 = q \approx 0.52$ and, to allow for detection of fainter attacks, $\delta_0 \approx 1.5$ (versus the estimated attack value $\delta \approx 1.9$).
} % End of ignore
We have implemented the score-based multi-cyclic SR and CUSUM procedures with the linear-quadratic score \eqref{eq:score}. When choosing the design parameters we assume the Gaussian model for pre-attack traffic, which agrees with the conclusions drawn above following the basic statistical analysis of the data. Thus, parameters $C_1, C_2$, and $C_3$ are chosen according to formulas~\eqref{eq:design_C} with $q_0 = q \approx 0.52$ and to allow for detection of fainter attacks $\delta_0 \approx 1.5$ (versus the estimated attack value $\delta \approx 1.9$). We set the detection thresholds $A \approx 1.9\times10^3$ and $h \approx 6.68$ so as to ensure the same level of $\ARL$ at approximately $500$ samples (i.e., $10$ sec) for both procedures. The thresholds are estimated using Monte Carlo simulations assuming the empirical pre-change distribution learned from the data. Specifically, we took $10^5$ samples from the empirical pre-change distribution and simulated the behavior of the respective detection statistics and procedures while adjusting the thresholds until observing the desired $\ARL$.

\begin{figure}[!htb]
    \centering
    \includegraphics[width=0.4\textwidth]{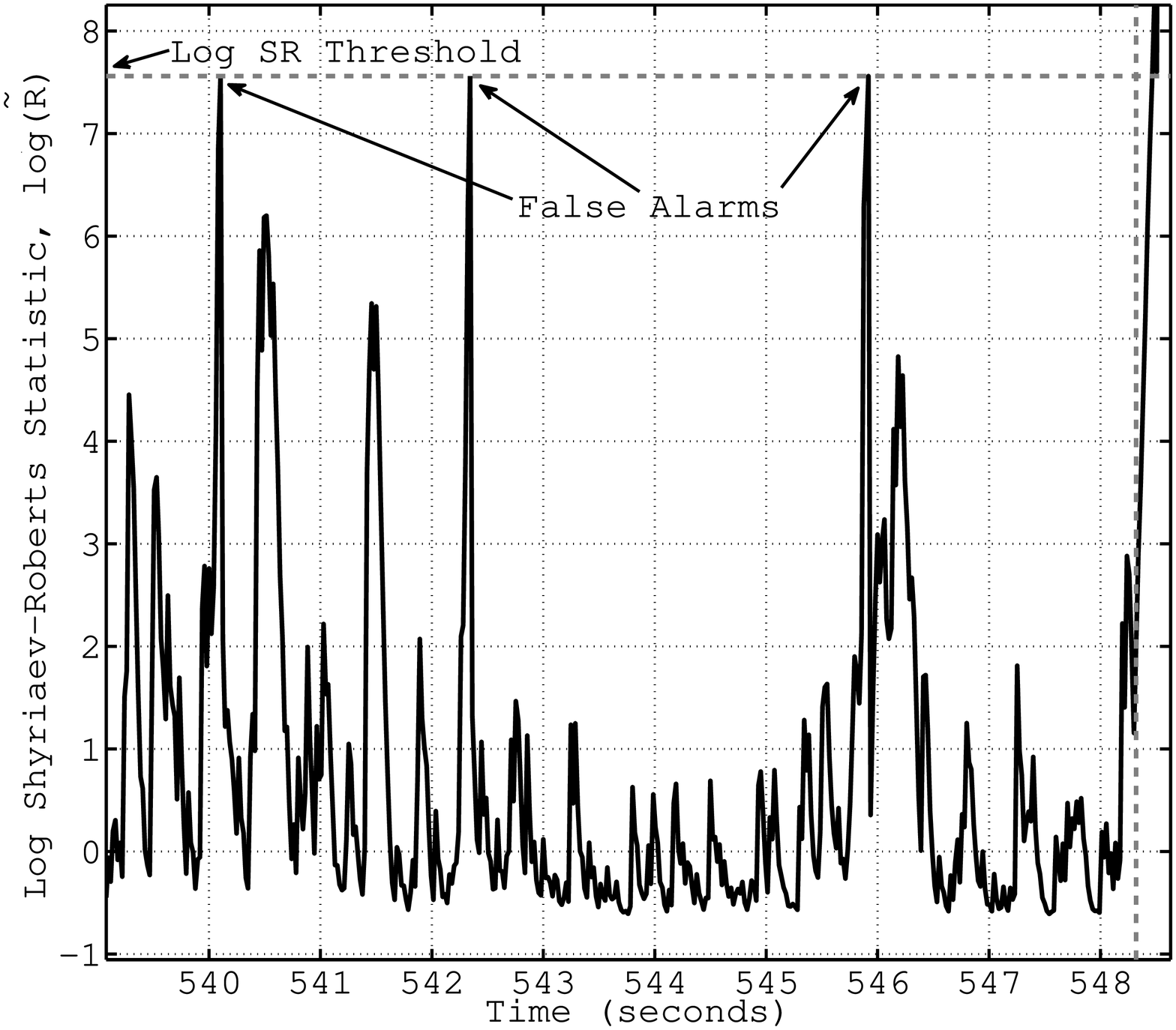}
    \caption{SYN flood attack: long run of the Shiryaev--Roberts procedure; logarithm of the SR statistic vs time.}
    \label{fig:SYN-false-alarms}
\end{figure}

\begin{figure}[!htb]
    \centering
    \subfigure[By the SR procedure]{
        \includegraphics[width=0.4\textwidth]{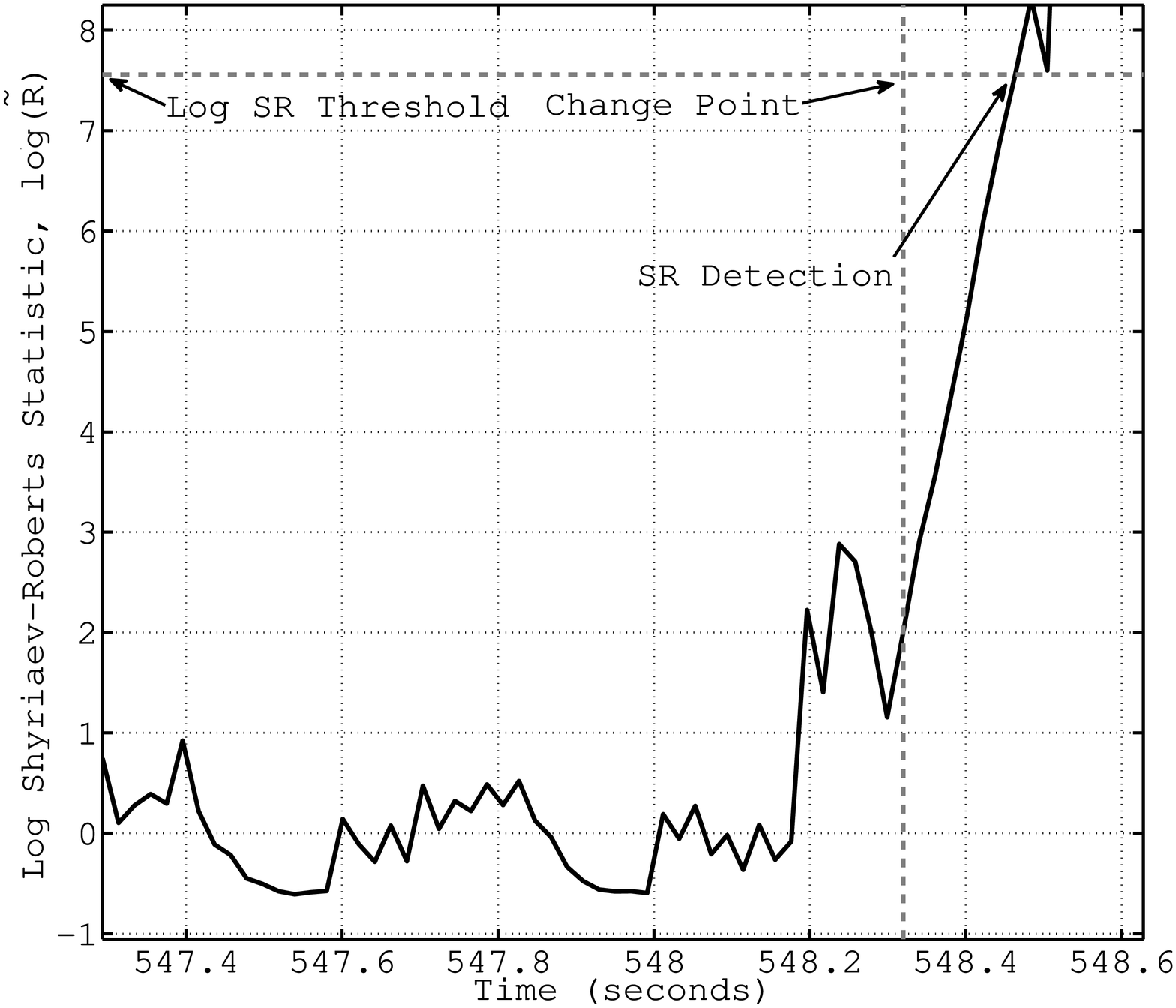}
        \label{fig:SYN-detection:SR}
    }
    \subfigure[By the CUSUM procedure]{
        \includegraphics[width=0.4\textwidth]{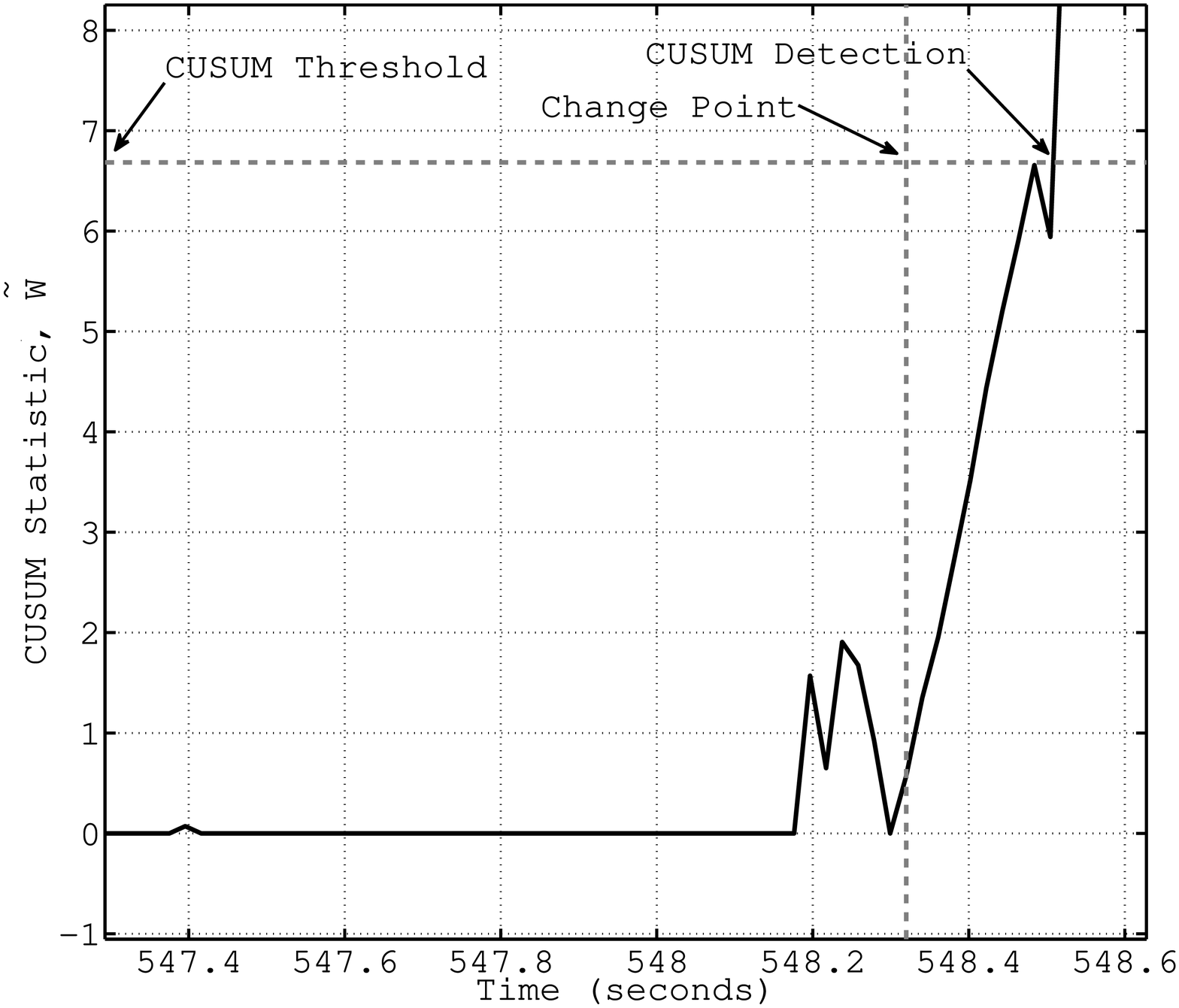}
        \label{fig:SYN-detection:CUSUM}
    }
    \caption{Detection of the SYN flood attack by the SR and CUSUM procedures.}
    \label{fig:SYN-detection}
\end{figure}

The detection process is illustrated in Figure~\ref{fig:SYN-false-alarms} and Figure~\ref{fig:SYN-detection}.   Figure~\ref{fig:SYN-false-alarms}  shows a relatively long run (taking into account the sampling rate $20$ msec) of the SR statistic with several false alarms and then the true detection of the attack with a very small detection delay (at the expense of raising many false alarms along the way). Recall that the whole idea of this paper is to set the detection thresholds low enough in order to detect attacks very quickly with minimal delays, which unavoidably leads to multiple false alarms prior to the attack starts. These false alarms should be  filtered by a specially designed algorithm, as has been suggested in \cite{Pollak+Tartakovsky:SS09} and will be  further discussed in Section~\ref{sec:further-discussion}.

Figure~\ref{fig:SYN-detection:SR} shows the behavior of the logarithm of the SR statistic shortly prior to the attack and right after the attack starts till its detection, which happens when the statistic crosses the threshold. Figure~\ref{fig:SYN-detection:CUSUM} shows the same for the CUSUM statistic. We see that both procedures successfully detect the attack with very small delays, though at the expense of raising false alarms along the way, as shown in Figure~\ref{fig:SYN-false-alarms} and discussed above. For both procedures we observed approximately $7$ false alarms per $1000$ samples. The detection delay for the repeated SR procedure is roughly $0.14$ seconds (or $7$ samples), and  for the CUSUM procedure the delay is about $0.21$ seconds (or $10$ samples). Thus, the SR procedure is better, as expected.
%-------------------------------------------------------------------------------------------------%
\section{Further Discussion}
\label{sec:further-discussion}

Since in real life legitimate traffic dominates, the idea of comparing various anomaly-based IDS-s using the multi-cyclic approach and the stationary average detection delay is a natural fit for cybersecurity applications.
However, it is worthwhile to remark on a possible way to enhance the potential of changepoint detection techniques as applied to cybersecurity. Any changepoint detection method is subject to the following drawback: instantaneous detection is not an option, unless the false alarm risk is high. Hence, though changepoint detection schemes are computationally inexpensive, in practice, employing one such scheme alone may not be a good idea, since it will be overflowed with false alarms. The simplest solution is to increase detection thresholds dramatically, but this will lead to an increase of the detection delay.

Here comes an interesting opportunity: What if one could combine changepoint detection techniques with others that offer very low false alarm rate, but are too heavy to use at line speeds? Do such synergistic anomaly detection systems exist, and how can they be integrated?

As an answer, consider complementing a changepoint detection-based anomaly detector with a flow-based signature IDS that examines the traffic's spectral profile. For an example of such signature-flow-based method, see, e.g.,~\cite{Christos6,He+etal:TR05,Hussain+etal:IEEE-INFOCOM06,He+etal:CN09}. The principal idea is to employ the Fourier transform to obtain the corresponding spectral characteristics of the passing traffic. This idea can be used in conjunction with the changepoint detection-based anomaly detector for both rejection of false alarms and confirmation of true detections. Higher computational complexity of the spectral-signature based detector is compensated by the preliminary changepoint anomaly based algorithm; the latter triggers the former only when there is a suspicion of an anomaly may be taking place in the network link of interest. For practical purposes the mean time between false alarms of the changepoint based anomaly IDS can be taken as small as a few seconds, as it was in the experiments presented in the previous section. We believe that such an alliance of the changepoint anomaly- and the spectral-signature-based detectors can significantly improve the whole system's overall performance reducing the false alarm rate to the minimum and at the same time guaranteeing very small detection delays.

%-------------------------------------------------------------------------------------------------%
\section{Conclusion}
\label{sec:conclusion}

We addressed the problem of rapid anomaly detection in computer network traffic. Approaching the problem statistically, namely, as that of sequential changepoint detection, we proposed a new anomaly detection method. The method is based on the multi-cyclic (repeated) Shiryaev--Roberts detection procedure where the likelihood ratio is replaced with the linear-quadratic score. This is done because in real-world network security applications both pre-attack and post-attack distributions are different from hypothesized distributions such as Gaussian or Poisson. Like many changepoint detection schemes, our method is also of practically no computational complexity and easy to implement. However, what distinguishes the SR procedure is its exact multi-cyclic optimality in a simple change detection problem where densities are known, a property that such techniques as the SPRT, the CUSUM chart, or the EWMA scheme lack. Hence, one may conjecture that the score-based SR detection algorithm is a better cyber ``watchdog''. To support this conjecture, we conducted a case study using a real SYN flood attack. The score-based multi-cyclic SR algorithm outperformed the multi-cyclic CUSUM procedure. Lastly, as a possible improvement of any changepoint detection-based anomaly detector, we proposed to complement the latter with a signature-based spectral IDS. This approach will allow to filter false alarms reducing the false alarm rate to a minimum and simultaneously guaranteeing prompt detection of real attacks.


\section*{Acknowledgments}

This work was supported in part by the U.S.\ National Science Foundation under grant CCF-0830419 and the U.S.\ Defense Threat Reduction Agency under grant HDTRA1-10-1-0086 at the University of Southern California, Department of Mathematics. The authors would like to thank Dr. Christos Papadopoulos (Colorado State University, Department of Computer Science) and Dr. John Heidemann (University of Southern California, Information Sciences Institute) for help with obtaining real data. The authors are also grateful to two anonymous referees whose comments helped to improve the paper.



\bibliographystyle{IEEEtran}
\bibliography{main,cybersecurity}

\end{document}